\documentclass[12pt]{iopart}
\usepackage{iopams}
\usepackage{setstack}
\usepackage{graphicx}

\newcommand{\cP}{\ensuremath{\mathcal{P}}}
\newcommand{\cT}{\ensuremath{\mathcal{T}}}
\newcommand{\half}{\mbox{$\textstyle{\frac{1}{2}}$}}
\newcommand{\quarter}{\mbox{$\textstyle{\frac{1}{4}}$}}
\newcommand{\threequarter}{\mbox{$\textstyle{\frac{3}{4}}$}}

\begin{document}

\rightline{preprint LA-UR-07-3496}
\title[Complexified Dynamical Systems]{Complexified Dynamical Systems}

\author[Bender, Holm, and Hook]{Carl~M~Bender$^*$\footnote{Permanent address:
Department of Physics, Washington University, St. Louis MO 63130, USA},
Darryl~D~Holm$^\dag$, and Daniel~W~Hook$^{\ddag}$}

\address{${}^*$Center for Nonlinear Studies, Los Alamos National Laboratory,
Los Alamos, NM 87545, USA\\{\footnotesize{\tt email: cmb@wustl.edu}}}
\vspace{.2cm}
\address{${}^{\dag}$Department of Mathematics, Imperial College, London SW7
2AZ, UK\\and\\
Computer and Computational Science, Los Alamos National Laboratory,\\
MS D413 Los Alamos, NM 87545, USA \\ {\footnotesize{\tt email: dholm@ic.ac.uk}}}
\vspace{.2cm}
\address{${}^{\ddag}$Blackett Laboratory, Imperial College, London SW7 2BZ, UK
\\{\footnotesize{\tt email: d.hook@imperial.ac.uk}}}

\begin{abstract}

Many dynamical systems, such as the Lotka-Volterra predator-prey
model and the Euler equations for the free rotation of a rigid body,
are $\cP\cT$ symmetric. The standard and well-known real solutions
to such dynamical systems constitute an infinitessimal subclass of
the full set of complex solutions.  This paper examines a subset of
the complex solutions that contains the real solutions, namely,
those having $\cP\cT$ symmetry. The condition of $\cP\cT$ symmetry
selects out complex solutions that are periodic.
\end{abstract}
\submitto{\JPA}

\section{Introductory Description of Classical $\cP\cT$ Symmetry}
\label{s1}
The differential equations that describe many classical dynamical systems are
$\cP\cT$ symmetric; that is, these equations are invariant under combined space
and time reflection. The equation for the simple pendulum \cite{R1}, the
Korteweg-de Vries and generalized Korteweg-de Vries equations \cite{R2}, the
Camassa-Holm equation \cite{R2,R3}, the Sine-Gordon equation \cite{R2}, the
Boussinesq equation \cite{R4}, and the classical equations \cite{R5,R6,R7,R8}
associated with some non-Hermitian quantum-mechanical systems
\cite{R9,R10,R11,R12} are all $\cP\cT$-symmetric.

In this paper we focus on Euler's differential equations, which govern the free
three-dimensional rotation of a rigid body about its center of mass. In
dimensionless form these equations may be written very simply as
\begin{equation}
\dot{L}_1=L_2L_3,\quad\dot{L}_2=-2L_1L_3,\quad\dot{L}_3=L_1L_2.
\label{e1}
\end{equation}
The three-dimensional classical dynamical system described by the
Euler equations has been studied in detail, the critical points are
known, the real solutions to these equations have been found, and
the structure of the phase-space trajectories is well understood
\cite{R13}. However, until now it has not been noticed that Euler's
equations, like the classical differential equations mentioned
above, are $\cP\cT$ symmetric. This paper examines the complex
$\cP\cT$-symmetric solutions to this system of equations. It also
explores this wider class of solutions for some other PT-symmetric
dynamical systems.

We begin by defining what is meant by $\cP\cT$ symmetry for classical physical
systems. We say that a real differential equation that describes classical
dynamics is $\cP\cT$ symmetric if it remains invariant under the combined
operations of parity reflection $\cP$, which changes the sign of all spatial
coordinates ${\bf x}$, and time reversal $\cT$, which changes the sign of the
time coordinate $t$. Furthermore, since the operation of time reversal in
quantum mechanics is associated with complex conjugation, we include complex
conjugation in the time-reversal operation when the differential equation is
complex. Thus, under $\cP\cT$ reflection we replace the dependent variable $f({
\bf x},t)$ of a differential equation by $f^*(-{\bf x},-t)$, where ${}^*$
represents complex conjugation. If $f^*(-{\bf x},-t)$ satisfies the same
differential equation as $f({\bf x},t)$, then the differential equation is
$\cP\cT$ symmetric.

For the special case of Hamiltonian systems of ordinary differential equations,
the solutions ${\bf x}(t)=[x_1(t),\,x_2(t),\,x_3(t),\,\ldots]$ and ${\bf p}(t)=
[p_1(t),\,p_2(t),\,p_3(t),\,\ldots]$ represent the dynamical coordinate and
momentum variables. Under $\cP\cT$ reflection ${\bf x}(t)$ is replaced by $-{\bf
x}^*(-t)$ and ${\bf p}(t)$ is replaced by ${\bf p}^*(-t)$. For the Euler
equations (\ref{e1}), the dependent variables $[L_1(t),L_2(t),L_3(t)]$
represent the {\it angular momentum} n the rotating frame of a rigid body and
therefore these variables transform under $\cP\cT$ reflection like spatial
coordinate variables: $[L_1(t),L_2(t),L_3(t)]\to[-L_1^*(-t),-L_2^*(-t),-L_3^*(
-t)]$. It is clear from this definition that the Euler equations (\ref{e1}) are
$\cP\cT$ symmetric. (We treat time $t$ as a {\it real} parameter; the case of
complex time is not considered in this paper.)

A solution ${\bf x}(t)$ to a system of dynamical differential equations is a
trajectory in coordinate space parameterized by time $t$. The $\cP\cT$
reflection of this curve, represented by $-{\bf x}^*(-t)$, is the mirror image
of the original curve ${\bf x}(t)$ reflected through the imaginary-$x$ axis. A
solution to a differential equation may or may not exhibit the symmetries of the
differential equation. An easy way to determine if the solution to a $\cP
\cT$-symmetric differential equation is itself $\cP\cT$ symmetric is to draw the
solution curve and to see if the curve is symmetric with respect to the
imaginary axis.
\smallskip

\begin{figure*}[t!]
\begin{center}
\includegraphics[bb=0 0 338 213,width=8cm]{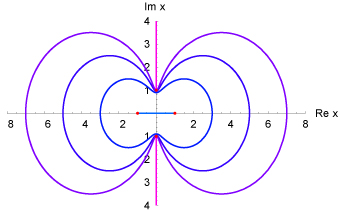}
\end{center}
\caption{Classical $\cP\cT$-symmetric trajectories in the complex-$x$ plane
representing the possible motions of a particle of energy $1$. This
motion is governed by the anharmonic-oscillator Hamiltonian $H=\half
p^2+x^4$. There is one real trajectory that oscillates between the
turning points at $x=\pm1$ and an infinite family of nested complex
trajectories that enclose the real turning points but lie inside the
imaginary turning points at $\pm i$. (The turning points are
indicated by dots.) Two other trajectories begin at the imaginary
turning points and drift off to infinity along the imaginary-$x$
axis. Apart from the trajectories beginning at $\pm i$, all
trajectories are closed and periodic. All closed orbits in this
figure have the same period $\sqrt{\pi/2}
\,\Gamma\left(\quarter\right)/\Gamma\left(\threequarter\right)=3.70815\ldots$.}
\label{f1}
\end{figure*}

\begin{figure*}[t!]
\begin{center}
\includegraphics[bb=0 0 443 276,width=9cm]{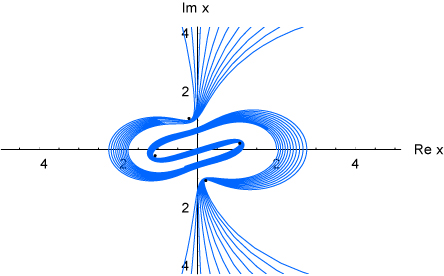}
\end{center}
\caption{A single non-$\cP\cT$-symmetric classical trajectory in the
complex-$x$ plane for a particle governed by the
anharmonic-oscillator Hamiltonian $H=\half p^2+x^4$. This trajectory
begins at $x=1$ and represents the complex path of a particle whose
energy $E=1+i$ is complex. The trajectory is not periodic because it
is not closed. The four turning points are indicated by dots.}
\label{f2}
\end{figure*}

\begin{itemize}
\footnotesize
\item[~~]
{\it Example 1: Complex classical trajectories for the anharmonic oscillator.}
The one-dimensional classical anharmonic oscillator Hamiltonian $H=\half p^2+
x^4$ gives Hamilton's equations
\begin{equation}
\dot{x}=p,\quad\dot{p}=-4x^3.
\label{e2}
\end{equation}
This dynamical system is $\cP\cT$ symmetric. There is one integral of the motion
(the energy is conserved), and thus the system can be reduced to a single
first-order equation:
\begin{eqnarray}
\half\dot{x}^2+x^4=E,
\label{e3}
\end{eqnarray}
where the integration constant $E$ may be complex. Note that even though
(\ref{e2}) is $\cP\cT$ symmetric, its first integral (\ref{e3}) need not respect
$\cP\cT$ symmetry. However, requiring that  (\ref{e3}) be $\cP\cT$ symmetric
translates into the physical condition that $E$ be real and allows us to
interpret $E$ as an energy.

$\quad$ Suppose first that $E$ is real and positive. We rescale $t$ and $x(t)$
so that $E=1$, and the four turning points in the complex-$x$ plane are located
at $\pm1$ and $\pm i$. The conventional real periodic motion of the system is
represented by a trajectory that lies on the real axis and oscillates between
the turning points at $\pm1$ (see Fig.~\ref{f1}). However, this real solution to
(\ref{e3}) is only one of an infinite number of possible complex trajectories
having the same energy. All but two of these trajectories lie outside the real
turning points and inside the complex turning points, as shown in Fig.~\ref{f1},
and they are closed and periodic. The remaining two trajectories each begin at
the turning points at $\pm i$ and run off to infinity in finite time along the
imaginary axis. Note that all of these trajectories are $\cP\cT$ symmetric; that
is, symmetric under reflections about the imaginary axis.

$\quad$ Next, suppose that $E$ is complex. For this case the resulting
trajectories are no longer $\cP\cT$ symmetric and are not closed and periodic.
In Fig.~\ref{f2} the trajectory for a particle whose energy is $E=1+i$ is
plotted. The trajectory begins at $x=1$ and does not close.
\end{itemize}
\smallskip
\normalsize

This anharmonic-oscillator example shows that the conventional real solutions to
the classical equations of motion (\ref{e2}) form a trivial lower-dimensional
subset of the class of complex $\cP\cT$-symmetric solutions shown in
Fig.~\ref{f1}. The $\cP\cT$-symmetric solutions in Fig.~\ref{f1} in turn
constitute a subset of the much larger class of all possible complex solutions
(see Fig.~\ref{f2}). However, in this paper we limit our attention to the $\cP
\cT$-symmetric classical solutions because these solutions have real energy.

The $\cP\cT$-symmetric solutions of the classical anharmonic oscillator are
special because they form {\it closed} and {\it periodic} orbits. Indeed, when
the classical particles exhibit $\cP\cT$-symmetric motion in the complex plane,
they are bound in a complex classical atom and cannot escape to infinity. (By
the term {\it complex atom} we mean a localized space-filling collection of
closed and periodic orbits.) If one quantizes this classical system using the
Bohr-Sommerfeld quantization condition $\oint dx\,p=\left(n+\half\right)\pi$,
one obtains the usual WKB approximation to the discrete energy levels of the
quantum anharmonic oscillator, independently of which closed classical orbit is
chosen as the integration path. (The Bohr-Sommerfeld quantization condition
cannot be applied to the {\it non}-$\cP\cT$-symmetric classical orbit in
Fig.~\ref{f2} because this orbit is not closed and periodic.) Since it is the
$\cP\cT$-symmetric classical orbits that give rise to the energies of the
associated quantum system, we regard the $\cP\cT$-symmetric orbits as physically
relevant. Thus, the correspondence principle establishes an association between
the family of $\cP\cT$-symmetric classical orbits and the quantum system.

The Bohr-Sommerfeld quantization condition requires that the classical orbit be
closed. Since the trajectory of a particle in the complex-$x$ plane cannot cross
itself, the condition of $\cP\cT$ symmetry is often strong enough to ensure that
the trajectory is closed. However, it is possible to have $\cP\cT$-symmetric
trajectories that are not closed~\cite{R7}. In such cases, the quantum
Hamiltonian has complex eigenvalues and is said to have a {\it broken} $\cP\cT$
symmetry. It is possible (though quite rare) to have {\it non}-$\cP
\cT$-symmetric trajectories that are closed and periodic, as the following
example shows.
\smallskip

\begin{figure*}[t!]
\begin{center}
\includegraphics[bb=0 0 339 211,width=9cm]{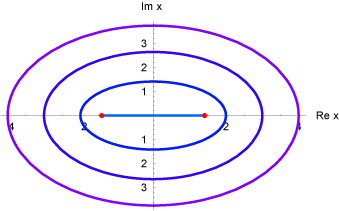}
\end{center}
\caption{Classical $\cP\cT$-symmetric trajectories in the
complex-$x$ plane for the harmonic-oscillator Hamiltonian $H=\half
p^2+\half x^2$. These trajectories are the complex paths
$x(t)=E\cos(t+A+iB)$ of a particle whose energy $E^2$ is real. The
trajectories shown are a family of nested ellipses with foci located
at the turning points denoted by dots at $x=\pm E\sqrt{2}$. We have
chosen $E$ to be real, so the turning points lie on the real-$x$
axis. (We could equally well have chosen $E$ to be imaginary, and in
this case the turning points would lie on the imaginary axis.) The
real line segment (a degenerate ellipse) connecting the turning
points is the conventional real periodic classical solution to the
harmonic oscillator. For $E^2=1$ the elliptical trajectories are
closed orbits, all having the same period $2\pi$.} \label{f3}
\end{figure*}

\begin{figure*}[t!]
\begin{center}
\includegraphics[bb=0 0 339 211,width=9cm]{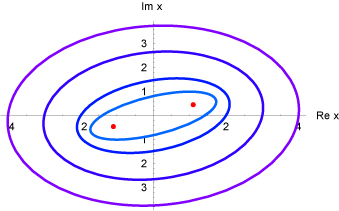}
\end{center}
\caption{Classical non-$\cP\cT$-symmetric trajectories in the
complex-$x$ plane for the harmonic-oscillator Hamiltonian $H=\half
p^2+\half x^2$. These trajectories are complex paths of the form
$x(t)=E\cos(t+A+iB)$ for a particle whose energy $E^2$ is complex.
The trajectories shown are not $\cP\cT$ symmetric, but they are
still closed and periodic. The two turning points are indicated by
dots.} \label{f4}
\end{figure*}

\begin{itemize}
\footnotesize
\item[~~]
{\it Example 2: Closed and periodic trajectories that are not $\cP\cT$
symmetric.} The one-dimensional classical harmonic oscillator Hamiltonian $H=
\half p^2+\half x^2$ has complex trajectories that are periodic, but not $\cP
\cT$ symmetric. For this Hamiltonian the equations of motion are
\begin{equation}
\dot{x}=p,\quad\dot{p}=-x.
\label{e4}
\end{equation}
These equations are $\cP\cT$ symmetric. There is one integral of the motion (the
energy is conserved), and thus this system can be reduced to the single
first-order equation
\begin{eqnarray}
\dot{x}^2+x^2=2E^2,
\label{e5}
\end{eqnarray}
where the energy $E^2$ is a constant that may be real or complex. The general
solution to (\ref{e5}) is $x(t)=E\cos(t+A+iB)$, where $A$ and $B$ are arbitrary
real constants. If $E$ is real, then
\begin{equation}
{\rm Re}\,x(t)=E\cos(t+A)\cosh(B),\quad {\rm Im}\,x(t)=E\sin(t+A)\sinh(B).
\label{e6}
\end{equation}
Thus, the graph of the trajectory $x(t)$ in the complex-$x$ plane is the ellipse
\begin{eqnarray}
[{\rm Re}\,x(t)]^2[E\cosh(B)]^{-2}+[{\rm Im}\,x(t)]^2[E\sinh(B)]^{-2}=1
\label{e7}
\end{eqnarray}
with semi-major and semi-minor axes $|E\cosh(B)|$ and $|E\sinh(B)|$. Ellipses of
the form in (\ref{e7}) are shown in Fig.~\ref{f3}. These ellipses are $\cP\cT$
symmetric; that is, they are symmetric with respect to reflections about the
imaginary axis. When $E$ is real (imaginary), the turning points at $\pm E$ lie
on the real (imaginary) axis. If the energy $E^2$ is complex, then the classical
trajectories are no longer $\cP\cT$ symmetric, as we can see in Fig.~\ref{f4}.
However, they are still closed and periodic.
\end{itemize}
\normalsize

The Lotka-Volterra equations provide a nice two-dimensional example of a $\cP
\cT$-symmetric dynamical system whose complex solutions are generally
nonperiodic but whose $\cP\cT$-symmetric complex solutions are periodic.

\begin{figure*}[t!]
\begin{center}
\includegraphics[bb=0 0 357 162,width=8cm]{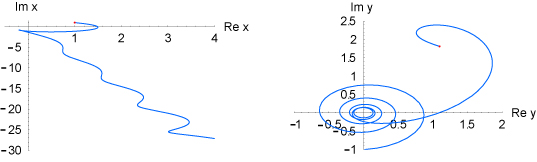}
\end{center}
\caption{Nonperiodic non-$\cP\cT$-symmetric complex solutions to the
Lotka-Volterra equations (\ref{e8}). For the initial conditions
$x(0)=1+i$ and $y(0)=1.09704+1.81173i$ the complex trajectories
$x(t)$ (left plot) and $y(t)$ (right plot) are clearly not periodic
and not $\cP\cT$ symmetric. (The initial conditions are indicated by
dots.)} \label{f6}
\end{figure*}

\smallskip
\begin{itemize}
\footnotesize
\item[~~]
{\it Example 3: $\cP\cT$-symmetric solutions to the Volterra equations}: If
we generalize slightly the definition of $\cP$ reflection to be $\cP:
\,(x,y)\to(y,x)$ \cite{R14}, then the Lotka-Volterra equations
\begin{equation}
\dot{x}=x-xy,\quad\dot{y}=-y+xy
\label{e8}
\end{equation}
become $\cP\cT$ symmetric. It is well known that the positive real solutions to
these equations are periodic they and provide a useful description of
predator-prey population dynamics, where $x(t)$ represents the population of the
prey species and $y(t)$ represents the population of the predator species. There
is one constant of the motion for the Lotka-Volterra equations:
\begin{equation}
x+y-\log(xy)=C.
\label{e9}
\end{equation}
For complex solutions $C$ is generally a complex constant, but for $\cP
\cT$-symmetric complex solutions $C$ must be real.

$\quad$ We first choose the set of initial conditions $x(0)=1+i$, $y(0)=2.11221-
0.403243i$ for which $C=2$ is real. The complex $\cP\cT$-symmetric solution,
which is plotted in Fig.~\ref{f5}, is closed and periodic. Next, we choose the
set of initial conditions $x(0)=1+i$, $y(0)=1.09704+1.81173i$ for which $C=1+i$
is complex. The complex solution, which is plotted in Fig.~\ref{f6}, is not $\cP
\cT$ symmetric and is not closed and periodic.

\begin{figure*}[t!]
\begin{center}
\includegraphics[bb=0 0 537 133,width=12cm]{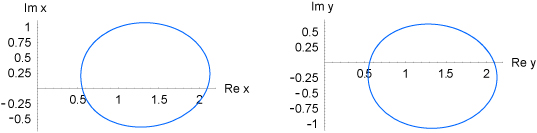}
\end{center}
\caption{Periodic $\cP\cT$-symmetric complex solutions to the
Lotka-Volterra equations (\ref{e8}). For the initial conditions
$x(0)=1+i$ and $y(0)=2.11221- 0.403243i$ the complex trajectories
$x(t)$ (left plot) and $y(t)$ (right plot) are shown. Observe that
the trajectories are periodic and $\cP\cT$ symmetric, where $\cP$
reflection interchanges $x$ and $y$ and $\cT$ reflection consists of
complex conjugation.} \label{f5}
\end{figure*}

\end{itemize}
\smallskip
\normalsize

Non-Hermitian quantum-mechanical Hamiltonians having unbroken $\cP\cT$ symmetry
are interesting because these operators have only real eigenvalues. Because of
the quantum-classical correspondence principle, some light might be shed on the
meaning of these quantum states by considering the $\cP\cT$-symmetric complex
solutions of the equations of classical mechanics. The first examples studied
have been the $\cP\cT$-symmetric complex solutions of the one-dimensional
nonlinear oscillator $H=p^2+x^2(ix)^\epsilon$ ($\epsilon\geq 0$)
\cite{R9,R10,R11,R12}. The Hermitian quantum version of the Hamiltonian that
describes rigid body rotation was treated in the doctoral thesis of Casimir
\cite{R15}. As far as we know, the $\cP\cT$-symmetric quantum version of this
problem has not yet been treated.

This paper examines the complex $\cP\cT$-symmetric solutions to Euler's
differential equations (\ref{e1}). In Sec.~\ref{s2} the Euler equations
(\ref{e1}) for free rigid-body rotation are derived and their complex solutions
are examined. Some concluding remarks are given in Sec.~\ref{s3}.

\section{Complex Angular Momentum Dynamics}
\label{s2}

The complex body-angular-momentum solutions ${\bf
L}(t)\in\mathbb{C}^3$ satisfy Euler's equations \cite{R16}
\begin{equation}
\mathbf{\dot{L}}=\frac{\partial C}{\partial\mathbf{L}}\times
\frac{\partial E}{\partial\mathbf{L}}, \label{e10}
\end{equation}
where $C$ and $E$ are conserved quadratic functions defined by
\begin{equation}
C(\mathbf{L})=\half\mathbf{L}\cdot\mathbf{L},\quad
E(\mathbf{L})=\half\mathbf{L}\cdot \mathbb{I}^{-1}\mathbf{L}.
\label{e11}
\end{equation}

Here, $\mathbb{I}^{-1}={\rm diag}(I_1^{-1},I_2^{-1},I_3^{-1})$ is
the inverse of the (real) moment-of-inertia tensor in principal-axis
coordinates. These equations are $\cP\cT$ symmetric; they are
invariant under spatial reflections of the angular momentum
components in the body $P:\,\mathbf{L}\to \mathbf{L}$ composed with
time reversal $T:\,\mathbf{L}\to-\mathbf{L}$.

We now make the simplifying choice $\mathbb{I}^{-1}\equiv{\rm
diag}(1,2,3)$, which reduces (\ref{e10}) to Euler's dynamical
equations in (\ref{e1}).  Equation (\ref{e1}) may be written
equivalently as
\begin{equation}
\mathbf{\dot{L}}=\mathbf{L}\times{\sf K}\mathbf{L} \qquad {\rm with}
\qquad {\sf K}\equiv{\rm diag}(-1,0,1).\label{e12}
\end{equation}

Since $\mathbf{L}$ is complex, we set $\mathbf{L}={\bf x}+i{\bf y}$
and obtain {\it four} conservation laws, the real and imaginary
parts of $C(\mathbf{L})=\half\mathbf{L}\cdot\mathbf{L}$ and
$H(\mathbf{L})=\half\mathbf{L}\cdot{\sf
K}\mathbf{L}=E(\mathbf{L})-2C(\mathbf{L})$, where

\begin{equation}
C(\mathbf{L})=\half\mathbf{x}\cdot\mathbf{x}
-\half\mathbf{y}\cdot\mathbf{y}+i\mathbf{x}\cdot\mathbf{y},\quad
H(\mathbf{L})=\half\mathbf{x}\cdot{\sf
K}\mathbf{x}-\half\mathbf{y}\cdot{\sf K}
\mathbf{y}+i\mathbf{x}\cdot{\sf K}\mathbf{y}. \label{e13}
\end{equation}

\begin{figure*}[t!]
\begin{center}
\includegraphics[bb=0 0 300 299,width=8cm]{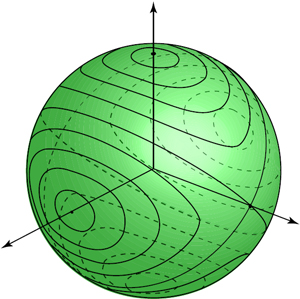}
\end{center}
\caption{Real phase-space trajectories for Euler's equations
(\ref{e1}). Choosing $C=\half$ in (\ref{e14}) limits these
trajectories to the surface of a three-dimensional sphere of radius
$1$. The critical points lie at $(\pm1,0,0)$, $(0,\pm1,0)$, and
$(0,0,\pm1)$. The trajectories for various values of $H$ in
(\ref{e14}) are shown.} \label{f7}
\end{figure*}

The solutions to Euler's equations that have been studied in the
past are the {\it real} solutions to (\ref{e12}), that is, the
solutions for which ${\bf y}= 0$. For this case the phase space is
three dimensional and the two conserved quantities are
\begin{equation}
C=\half\left(x_1^2+x_2^2+x_3^2\right),\quad H=\half x_3^2-\half
x_1^2. \label{e14}
\end{equation}
If we take $C=\half$, then the phase-space trajectories are
constrained to a sphere of radius $1$. There are six critical points
located at $(\pm1,0,0)$, $(0 ,\pm1,0)$, and $(0,0,\pm1)$. The
trajectories for various values of $H$ are shown in Fig.~\ref{f7}.
These are the conventional trajectories that are discussed in
standard textbooks on dynamical systems \cite{R13}. (When $H=0$, the
resulting equation is a first integral of the simple pendulum
problem \cite{R1}.)

Let us now examine the complex $\cP\cT$-symmetric solutions to
Euler's equations (\ref{e12}). For this case phase space is six
dimensional, which is difficult to visualize. However, the
requirement of $\cP\cT$ symmetry implies that the constants of
motion $C$ and $H$ in (\ref{e13}) are real. The vanishing of the
imaginary parts of $C$ and $H$ gives the two equations:
\begin{equation}
\mathbf{x}\cdot\mathbf{y}=0,\quad\mathbf{x}\cdot{\sf K}\mathbf{y}=0.
\label{e15}
\end{equation}
These two bilinear constraints may be used to eliminate the
$\mathbf{y}$ terms in the complex equations (\ref{e12}). When this
elimination is performed one obtains the following real equations
for $\mathbf{x}$ on the $\cP\cT$ constraint manifolds (\ref{e15}):
\begin{equation}
\mathbf{\dot{x}}=\mathbf{x}\times{\sf
K}\mathbf{x}+M(\mathbf{x})\,\mathbf{x}. \label{e16}
\end{equation}
Here, the scalar function $M=PN/D$, where the functions $P$, $N$,
and $D$ are given by
\begin{eqnarray}
P(\mathbf{x})&=&2x_1x_2x_3,\qquad N(\mathbf{x})=x_1^2+x_2^2+x_3^2-1,\nonumber\\
D(\mathbf{x})&=&\left|{\rm Re}\,\left(\frac{\partial
C}{\partial\mathbf{L}} \times\frac{\partial
H}{\partial\mathbf{L}}\right)\right|^2
=x_1^2x_2^2+x_2^2x_3^2+4x_1^2x_3^2. \label{e17}
\end{eqnarray}
(This form may be helpful in thinking about the PT-symmetric quantum
spin problem.)

\begin{figure*}[t!]
\begin{center}
\includegraphics[bb=0 0 700 481,width=15cm]{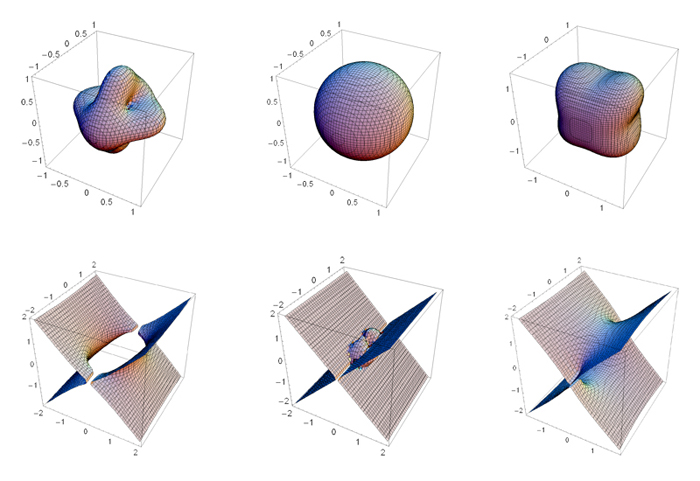}
\end{center}
\caption{Level surfaces in $(x_1,x_2,x_3)$ space for the two
constants of the motion, $A$ and $B$ in (\ref{e19}). The top row
shows the level surfaces corresponding to $A=-1$, $0$, and $1$ and
the bottom row shows the level surfaces for $B=-1$, $0$, and $1$.}
\label{f8}
\end{figure*}

The system (\ref{e16}) has nonzero divergence, so it cannot be
Hamiltonian even though it arises from constraining a Hamiltonian
system. Nonetheless, the system has two {\it additional} real
conservation laws, and it reduces to the integrable form
\begin{eqnarray}
\dot{x}_1=x_2x_3\left(1+2x_1^2N/D\right),&&\quad\dot{x}_2=-2x_1x_3\left(
1-x_2^2N/D\right),\nonumber\\
&&\!\!\!\!\!\!\!\!\!\!\!\!\!\!\!\!\!\!\!\!\!\!\!
\dot{x}_3=x_1x_2\left(1+2x_3^2N/D\right) \label{e18}
\end{eqnarray}
on level sets of two conserved quantities:
\begin{equation}
A=\frac{(N+1)^2N}{D},\quad
B=\frac{x_1^2-x_3^2}{D}\left(2x_2^2x_3^2+4x_1^2x_3^2+
x_2^4+2x_1^2x_2^2-x_2^2\right). \label{e19}
\end{equation}
In Fig.~\ref{f8} the level surfaces in $(x_1,x_2,x_3)$ space are
displayed for three values of $A$ and for three values of $B$.

\begin{figure*}[t!]
\begin{center}
\includegraphics[bb=0 0 400 434,width=8cm]{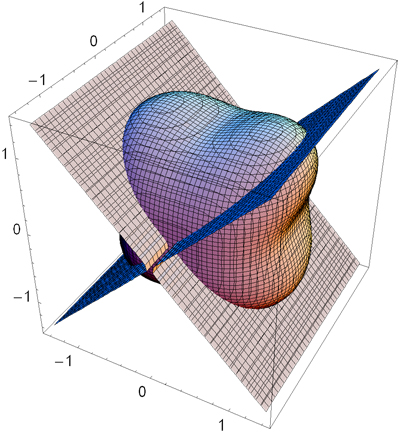}
\end{center}
\caption{Superposition of two level surfaces in Fig.~\ref{f8}
corresponding to $A=1$ and $B=0$. The trajectory, which is the
intersection of these two level surfaces, is a closed
butterfly-shaped curve in $(x_1,x_2,x_3)$ space.} \label{f9}
\end{figure*}

\begin{figure*}[t!]
\begin{center}
\includegraphics[bb=0 0 510 577,width=8cm]{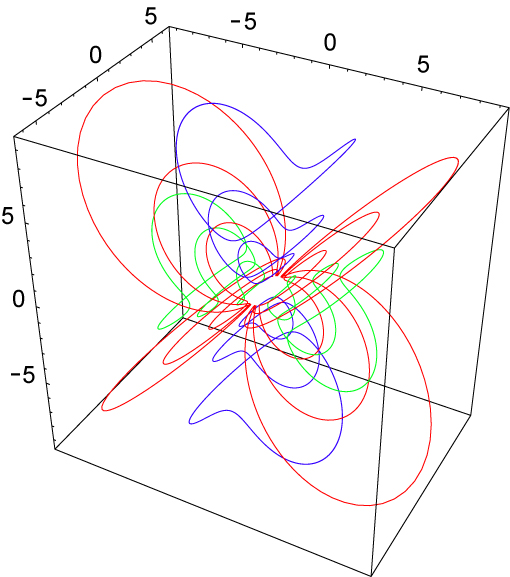}
\end{center}
\caption{Closed and periodic butterfly-shaped trajectories plotted
in $(x_1,x_2,x_3)$ space. At the center of the figure there are no
trajectories indicated. All curves shown in this figure lie outside
the unit ball centered at the origin. Trajectories on the unit ball
are shown in Fig.~\ref{f7}. The trajectories that lie inside the
unit ball are shown in Fig.~\ref{f11}. Trajectories outside the unit
ball never cross to the interior of the unit ball.} \label{f10}
\end{figure*}

The $\cP\cT$-symmetric trajectories in $(x_1,x_2,x_3)$ space are
characterized by the values of $A$ and $B$ in (\ref{e19}) and these
trajectories are precisely the intersections of the level surfaces
shown in the top and bottom rows of Fig.~\ref{f8}. For example, in
Fig.~\ref{f9} the level surfaces corresponding to $A=1$ and $B=0$
are superposed. One can see that the intersection of these surfaces
is a pair of closed butterfly-shaped curves in $(x_1,x_2,x_3)$
space. In Fig.~\ref{f10} many such butterfly-shaped trajectories are
shown, and all such trajectories are closed and periodic. All of
these trajectories lie outside a unit ball centered at the origin.

\begin{figure*}[t!]
\begin{center}
\includegraphics[bb=0 0 517 558,width=8cm]{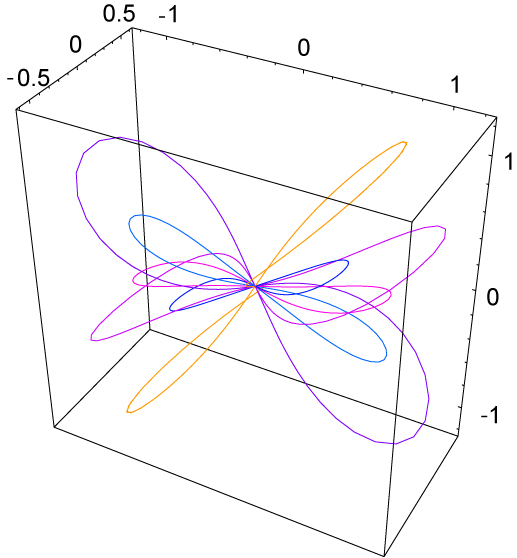}
\end{center}
\caption{Closed and periodic
butterfly-shaped trajectories that lie inside the unit ball centered
at the origin. Every trajectory passes through the origin.}
\label{f11}
\end{figure*}

In Fig.~\ref{f11} we display the $\cP\cT$-symmetric trajectories
that lie inside the unit ball centered at the origin. All of these
trajectories pass through the origin. However, these trajectories
are periodic and do not stop at the origin. While one can see from
(\ref{e1}) that the origin is a critical point for real
trajectories, it is clear from (\ref{e17}) and (\ref{e18}) that the
origin in $(x_1,x_2,x_3)$ space is {\it not} a critical point.

\section{Conclusions}
\label{s3}

\begin{figure*}[t!]
\begin{center}
\includegraphics[bb=0 0 581 444,width=8cm]{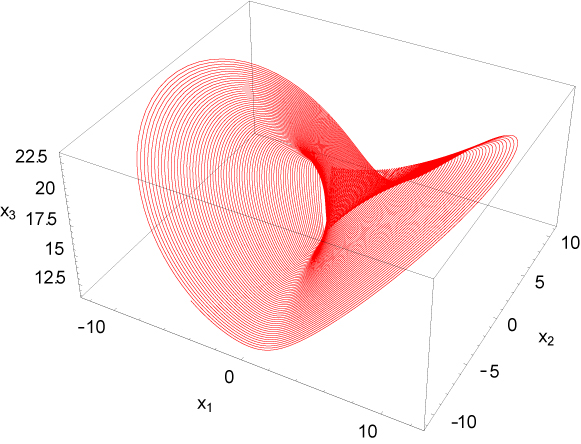}
\end{center}
\caption{A single nonperiodic butterfly-shaped open trajectory that
arises when the constants of motion $H=1+i$ and $C=1+i$ in
(\ref{e13}) are complex. The trajectory is not $\cP\cT$ symmetric
and not closed. Rather, it spirals out to infinity.} \label{f12}
\end{figure*}

In this paper we have used the Euler equations for the free rotation of a rigid
body about its center of mass to illustrate the following picture: Given a real
dynamical system of differential equations, the real solutions form a tiny
subset of the rich and interesting class of complex solutions. If this system is
$\cP\cT$ symmetric, then some of the complex solutions will themselves be $\cP
\cT$ symmetric. The $\cP\cT$-symmetric solutions are characterized by having
{\it real} constants of the motion, such as the energy, and thus we view these
solutions as being physical. The $\cP\cT$-symmetric trajectories are different
from the other complex trajectories in that they are closed and periodic. For
example, if we take the constants of the motion in (\ref{e13}) to be complex,
$H=1+i$ and $C=1+i$, then we see in Fig.~\ref{f12} that while the complex
trajectories resemble the butterfly-shaped orbits shown in Fig.~\ref{f10}, they
are {\it open} rather than closed and periodic orbits. Our results indicate that
the following three statements are equivalent: (i) A complex rotating-rigid-body
solution is $\cP\cT$ symmetric; (ii) its constants of motion are real; (iii) it
is periodic (except for possible heteroclinic cycles).

It is crucial that the trajectories of a classical dynamical system
be multiply periodic because only then do the methods of
Bohr-Sommerfeld quantization apply. In conventional quantum
mechanics only the real phase-space trajectories are considered, and
these trajectories terminate at the turning points. On the other
side of the turning points is the so-called ``classically-forbidden
region." We take a broader view and argue that the set of all
$\cP\cT$-symmetric classical orbits constitutes a complex atom. Any
one of these orbits can be used to determine the energy levels of
the associated quantum system because these orbits are closed and
thus Cauchy's theorem implies that the energy levels so obtained
will be unique. Furthermore, the reality of the energy levels, which
is a crucial property of a physical quantum system, is a consequence
of the $\cP \cT$ symmetry of the underlying classical system.

\vspace{0.5cm}
\footnotesize
\noindent
We thank J.~Bender for pointing out Ref.~\cite{R14} and DWH thanks M.~Dixon for
helpful conversations. CMB thanks the Mathematics Department at Imperial
College, London, for its hospitality. As an Ulam Scholar, CMB receives financial
support from the Center for Nonlinear Studies at the Los Alamos National
Laboratory. CMB is also supported by a grant from the U.S. Department of Energy.
The work of DDH is supported by the Royal Society of London and by the
U.S.~Department of Energy Office of Science Applied Mathematical Research.

\end{document}